\documentclass[a4paper,amsmath,amssymb,prl,floatfix,footinbib,twocolumn,preprintnumbers]{revtex4-1}

\usepackage[dvips]{graphicx}
\usepackage{dcolumn}
\usepackage{bm}
\usepackage{dsfont}
\usepackage{color}
\usepackage[normalem]{ulem}
\usepackage{url}
\usepackage[colorlinks=true ,citecolor=blue]{hyperref}
\usepackage{amsmath,amssymb}

\def\be{\begin{equation}}
\def\ee{\end{equation}}
\def\bea{\begin{eqnarray}}
\def\eea{\end{eqnarray}}
\def\ba{\begin{array}}
\def\ea{\end{array}}
\def\pa{\partial}

\def\Om{\Omega}
\def\om{\omega}

\def\nn{\nonumber}
\def\ket{\rangle}

\def\b{\mathbf}

\def\f{\frac}

\def\ra{\rightarrow}
\def\rm{\mathrm}

\newcommand{\mv}[1]{\langle #1\rangle}

\begin{document}

\title{Particle-hole character of the Higgs and Goldstone modes \\ in
  strongly-interacting lattice bosons}

\author{M. Di Liberto$^1$}
\email{mar.diliberto@gmail.com}
\author{A. Recati$^{1,2}$}
\email{alessio.recati@unitn.it}
\author{N. Trivedi$^3$}
\author{I. Carusotto$^1$}
\author{C. Menotti$^1$}
\affiliation{ $^1$INO-CNR BEC Center and Dipartimento di Fisica,
  Universit\`a di Trento, 38123 Povo, Italy
  \\ $^{2}$ Arnold Sommerfeld Center for Theoretical Physics,
  Ludwig-Maximilians-Universit\"at M\"unchen, 80333 M\"unchen,
  Germany \\ $^{3}$ Department of
  Physics, The Ohio State University, Columbus, OH 43210, USA}

\date{\today}

\begin{abstract}

We study the low-energy excitations of the Bose-Hubbard model in the
strongly-interacting superfluid phase using a Gutzwiller approach and 
extract the single-particle and single-hole excitation
amplitudes for each mode. We report emergent mode-dependent
particle-hole symmetry on specific arc-shaped lines in the phase diagram 
connecting the well-known Lorentz-invariant limits of the Bose-Hubbard model. By tracking the 
in-phase particle-hole symmetric oscillations of
the order parameter, we provide an answer to the long-standing question about the fate of the pure
amplitude Higgs mode away from the integer-density critical point. Furthermore, we 
point out that out-of-phase oscillations are responsible for a full
suppression of the condensate density oscillations of the gapless
Goldstone mode. Possible detection protocols are also discussed.

\end{abstract}


\maketitle


\emph{Introduction.} 
Ultra-cold atoms in optical lattices provide an ideal platform 
to explore the properties of strongly-interacting lattice systems. 
A prominent example of their capability to
reproduce prototypical lattice Hamiltonians is given by the experimental 
realization of the Bose-Hubbard model \cite{Fisher1989,Jaksch1998,Greiner2002,Bloch2008}.
Indeed, the superfluid to Mott insulator
transition has been characterized at a very high level of accuracy,
exhibiting an excellent agreement between experiments and theoretical
predictions at zero \cite{Greiner2002} and finite temperature
\cite{Kato2008,Trotzky2010}.  The ground-state properties of the Bose-Hubbard
model have been thoroughly investigated through time-of-flight imaging
\cite{Greiner2002}, measure of noise correlations \cite{Folling2005}
and single-site microscopy \cite{Bakr2009}. Excitations have been also
addressed through tilting of the lattice \cite{Greiner2002}, Bragg
spectroscopy \cite{Ernst2009} and lattice depth modulation
\cite{Stoferle2004}.

A very intriguing regime of the Bose-Hubbard model is the strongly-interacting superfluid phase.
A clear distinctive feature with
respect to a weakly-interacting superfluid is provided by 
the strong particle or hole character of the phonons of the superfluid
close to the Mott lobes.
In contrast to the weakly-interacting limit, where the gapless Goldstone mode
exhausts all of the spectral weight, a further signature of strong
correlations is the existence of gapped modes
\cite{SachdevBook,Ohashi2006,Menotti2008,Pekker2015}.
The first gapped mode has been recently observed in the short
wavelength limit using Bragg spectroscopy \cite{Bissbort2011} and in
the large wavelength limit using lattice modulation \cite{Endres2012}.
When the first gapped mode consists of a pure amplitude oscillation of
the superfluid order parameter \cite{Sachdev1999,Zwerger2004,Altman2002,Podolsky2011,Pollet2012,Podolsky2012,Gazit2013}, it is granted the label of {\it Higgs
 mode}, in analogy with the Higgs
boson in particle physics \cite{Pekker2015}.

A pure amplitude mode, decoupled from the phononic phase mode, has
been predicted to exist when the Bose-Hubbard model is effectively
described by a relativistic $O(2)$ field theory, since an effective
particle-hole symmetry ensures Lorentz invariance and the resulting
decoupling of phase and amplitude degrees of freedom
\cite{SachdevBook,Altman2002,Huber2007,Huber2008}. This $O(2)$ theory
describes both the vicinity of the critical point of the superfluid to
Mott transition at integer filling in dimensions $d\geq 2$ and
hard-core bosons at half-integer filling \cite{Lindner2010}.

An important issue regards the fate of the Higgs mode away from
criticality and towards the weakly-interacting regime ~\cite{Pekker2015}.
To the best of our knowledge, no clear answer about this question has been
provided yet.

In this Letter,
we find an emergent particle-hole symmetry for the first gapped mode
on a curve connecting two Lorentz-invariant points of the model,
starting from the tip of the insulating lobes.
This result relies on higher-energy excitations and provides an answer to
the long-standing debate about the conditions of existence of a
pure-amplitude {\it Higgs} mode in the Bose-Hubbard model away from criticality. 
Moreover, we show that a distinct particle-hole
symmetry condition for the gapless Goldstone mode produces 
a suppression of the condensate density oscillations in proximity 
of the Mott lobes and, specifically,
in correspondence to the boundary between particle and hole
superfluidity. We speculate that such a suppression may be responsible for an increase
in the critical temperature of the normal to superfluid transition.

\begin{figure*}[!tbp]
\includegraphics[width=1\textwidth]{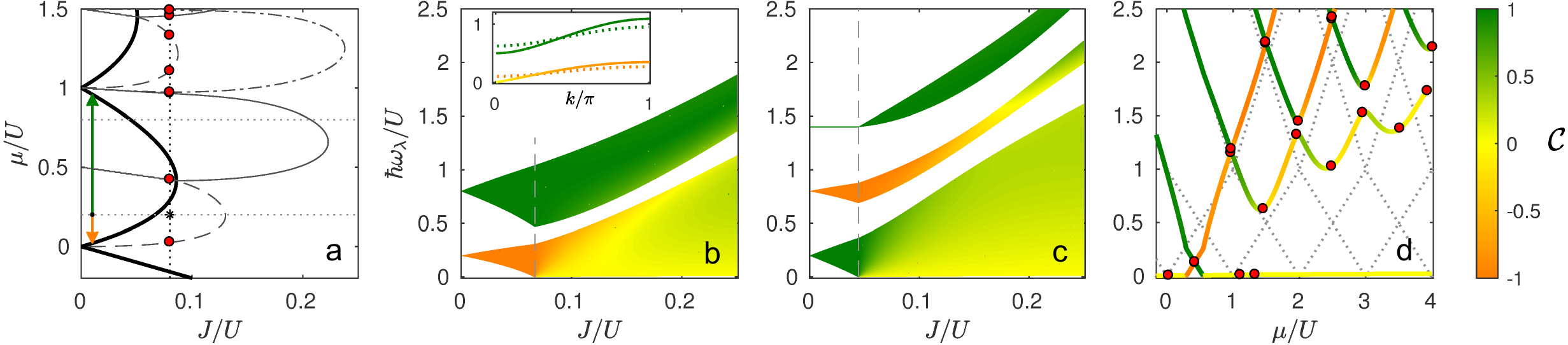}
\caption{(a) Mean-field phase diagram of the Bose-Hubbard model: thick
  lines are the Mott insulator phase boundaries. For simplicity, we consider
  a one-dimensional lattice in all the figures. Single-particle
  (-hole) excitations in the Mott phase are indicated by the green
  (orange) arrows. Dashed, solid and dashed-dotted grey arcs
  indicate the condition of particle-hole symmetry $\mathcal C = 0$ at $k\approx 0$ for the Goldstone,
  first and second gapped modes, respectively.  (b,c) Lowest bands as a function of $J/U$ for
  $\mu/U=0.2$ (b) and $\mu/U=0.8$ (c) (see horizontal dotted lines in (a)). 
  The vertical dashed line
  indicates the phase transition. Inset in (b): Excitation spectrum as
  a function of $k/\pi$ in the strongly interacting superfluid at
  $J/U=0.08$ and $\mu/U=0.2$ (star in (a)) compared with the
  excitation spectrum in the Mott phase at $J/U=0.04$ and $\mu/U=0.2$
  (dotted lines). In all figures, line color indicates the value of
  ${\cal C}$, quantifing the particle-hole character for each
  mode. Particle-hole symmetry is found when ${\cal C}=0$. 
   (d) Grey dotted lines
  of slope $\pm m \mu/U$ are the $m$-holes (or $m$-particles)
  excitation energies in the Mott phase at $J=0$.  Thick lines:
  Excitation energies $\hbar \omega_{k,\lambda}/U$ for modes
  $\lambda=1\dots 4$ at $k =\pi/100 $ along the vertical dotted
  line in (a), namely as a function of $\mu/U$ for $J/U=0.08$.  The
  points of particle-hole symmetry are highlighted by the red dots
  (see also (a)). }
\label{fig:spectrum}
\end{figure*}


\emph{Model and theory.}
The Bose-Hubbard model for a uniform lattice reads
\be
H = - J \sum_{\mv{i,j}}\left( a^\dag_i a^{}_j + \rm{H.c.} \right) + \f{U}{2}\sum_i n_i (n_i-1) - \mu \sum_i n_i \,,
\ee
where $J$ is the hopping amplitude, $U$ the on-site interaction and
$\mu$ the chemical potential. We study the excitations of the system
by means of a time-dependent Gutzwiller ansatz $ |\psi\ket = \prod_i 
\sum_n c_{i,n}(t)|n\ket_i $, where the coefficients $c_{i,n}(t)$ satisfy the equations of motion
obtained from the Lagrangian $L [c,c^*] \equiv  
i\hbar \sum_{i,n} \, c^{*}_{i,n} \pa_t c_{i,n} - \mv{H} $.  

We define $c_{i,n}(t) = \left[\bar c_{n} +\delta c_{i,n}(t)
  \right]e^{-i\om_0 t}$
\cite{StringariBook,Krutitsky2011,Krutitsky2016}, where $\bar c_{n}$
are the ground state parameters, $\om_0$ describes the time dependence
at equilibrium, and $\delta c_{i,n}(t)$ are the small oscillations
with respect to the equilibrium configuration.
Linearizing the equations of motion with respect to $\delta
c_{i,n}(t)$ and introducing the Ansatz $\delta c_{i,n}(t) = u_{\b{k},n}
e^{i(\b{k}\cdot \b{r}_i-\om_{\b k} t)} + v_{\b{k},n} e^{-i(\b{k}\cdot
  \b{r}_i-\om_{\b k} t)} $, one obtains Bogoliubov-like equations for 
the coefficient $u_{\b{k},n}$ and $v_{\b{k},n}$, which can be chosen to be real. 
To describe the excitations above the ground state, we select the solutions at positive 
energy $\omega_{{\bf k},\lambda}>0$, where $\lambda=1,2,\dots$ identifies the
different branches of the spectrum. The corresponding eigenvectors
satisfy $\vec u_{\b k, \lambda} \cdot \vec u_{\b k, \lambda'} - 
\vec v_{\bf k, \lambda} \cdot \vec v_{\b k, \lambda'} = 
\varepsilon \,\delta_{\lambda,
  \lambda'}$, with $\varepsilon>0$.  For practical convenience,
  we take $\varepsilon=1$. 

Given a certain observable $A$, an excitation $\lambda$ produces a perturbation
with respect to the ground state value $\delta A_\lambda = \mv{A}_{\lambda} - \bar{A}$,
which we consider up to linear order in $\delta c_{i,n}$.
For the order parameter $\psi_i = \mv{a_i}$, this reads
\bea
\delta \psi_{i,\lambda}
&=& \mathcal U_{\b k,\lambda} e^{i(\b{k}\cdot
  \b{r}_i-\om_{\b k, \lambda} t)} + \mathcal V_{\b k, \lambda}
e^{-i(\b{k}\cdot \b{r}_i-\om_{\b k, \lambda} t)}, \label{dphis}
\eea
where 
\bea
\mathcal U_{\b k, \lambda} &=& \sum_n \sqrt{n+1}\left(\bar c_{n}
u^{(\lambda)}_{\b{k},n+1}+\bar c_{n+1}
v^{(\lambda)}_{\b{k},n}\right)\,,\nn\\
\mathcal V_{\b k, \lambda} &=&
\sum_n \sqrt{n+1}\left(\bar c_{n+1} u^{(\lambda)}_{\b{k},n}+\bar c_{n}
v^{(\lambda)}_{\b{k},n+1}\right)\,.
\eea
The quantities $|\mathcal U_{\b k, \lambda}|^2$ and $|\mathcal V_{\b k,
  \lambda}|^2$ are the quasi-particle and quasi-hole 
  excitation strengths for mode $\lambda$, respectively
\cite{Krutitsky2011,Krutitsky2016}.
%


\emph{Particle-hole symmetry.} For each mode and momentum, 
we define particle-hole symmetry the condition 
$|\mathcal U_{\b k, \lambda}|=|\mathcal V_{\b k, \lambda}|$,
identified by the zeros of the function ${\cal C}=(|\mathcal U_{\b
  k,\lambda}| - |\mathcal V_{\b k,\lambda}|)/(|\mathcal U_{\b k,\lambda}| + |\mathcal 
V_{\b k,\lambda}|)$ (see Fig.~\ref{fig:spectrum}) \footnote{Usually, particle-hole
symmetry refers to the invariance of the Hamiltonian under a transfomation
of the form $a \ra a^\dag$. In this work, we use the term particle-hole symmetry 
to describe excitations with equal particle and hole strengths.}.
To understand the existence of lines of particle-hole symmetry, it
is helpful to recall how the excitations in the Mott phase evolve into the
phononic and gapped modes of the strongly-interacting superfluid
\cite{Menotti2008}.  This simple observation discloses properties of
the excitation modes that are crucial to the aim of the present work.

In the weakly-interacting Bogoliubov regime, phonons present a
strong particle and hole admixture. 
In contrast, close to the Mott lobes, the phononic excitations of the 
strongly-interacting superfluid inherit the pure particle or hole character
of the Mott excitation that becomes gapless at the transition 
(see Fig.~\ref{fig:spectrum}(a)). For negative (positive)
doping with respect to integer filling, phononic excitations of the
strongly-interacting superfluid appear with $|\mathcal V_{\b k,1}| \gg
|\mathcal U_{\b k,1}|$ ($|\mathcal U_{\b k,1}| \gg |\mathcal V_{\b
  k,1}|$), indicating dominant hole (particle) character  (see Fig.~\ref{fig:spectrum}(b-c), respectively).  
  At negative (positive) doping, the second lowest Mott excitation is gapped at the
transition and is transformed into the first gapped mode of the
superfluid phase, which conversely has particle (hole) character
$|\mathcal U_{\b k,2}| \gg |\mathcal V_{\b k,2}|$ ($|\mathcal V_{\b
  k,2}| \gg |\mathcal U_{\b k,2}|$) (see Fig.~\ref{fig:spectrum}(b-c), respectively).

It is instructive to realize that also higher excited modes inherit
their particle-hole character from underlying pure $m$-particle and
$m$-hole excitations (see Fig.~\ref{fig:spectrum}(d)). 
The energy crossing between such excitations turn into anti-crossings
due to the coupling introduced by a non-vanishing order parameter in
the superfluid phase.  The dominant particle or hole character from
the underlying modes is retained, except in the vicinity of the
anti-crossing points, where hybridization 
leads to a point of perfect particle-hole symmetry for each mode.
In the weakly-interacting regime all excitations become
particle-dominated. Hence, the regions of dominant
hole-character are confined in the strongly-interacting regime and bounded by
a line of perfect particle-hole symmetry (${\cal C}=0$, see grey curves in
Fig.~\ref{fig:spectrum}(a)).
This picture highlights the role played by energetically-close
excitations in determining the particle-hole symmetry condition for
the different modes.
In particular, the idea of particle-hole symmetry arising close to
energy level crossings explains why particle-hole symmetry is
recovered for all modes in the vicinity of the tip of the lobes ($\mu/U$
close to half-integer values) and at very small $J/U$ and half-integer filling ($\mu/U$
close to integer values) (see Fig.~\ref{fig:spectrum}(a,d)).

In the following, we are going to discuss how in-phase and out-of phase oscillations
of the order parameter (namely the relative sign of $\,\mathcal U_{\b k, \lambda}$ 
and $\mathcal V_{\b k, \lambda}$) at the
particle-hole symmetry condition determine profoundly-different physical properties of the
two lowest-lying excitations \footnote{We have numerically observed that all
  odd (Goldstone and further) excitation modes present arc-shaped lines 
  of the phase diagram where 
  $\mathcal R_{\b k, \lambda} = 0$, while all even (``Higgs'' and further) modes
  present arc-shaped lines where $\mathcal I_{\b k, \lambda} = 0$.}.


\begin{figure}[!tbp]
\includegraphics[width=1\columnwidth]{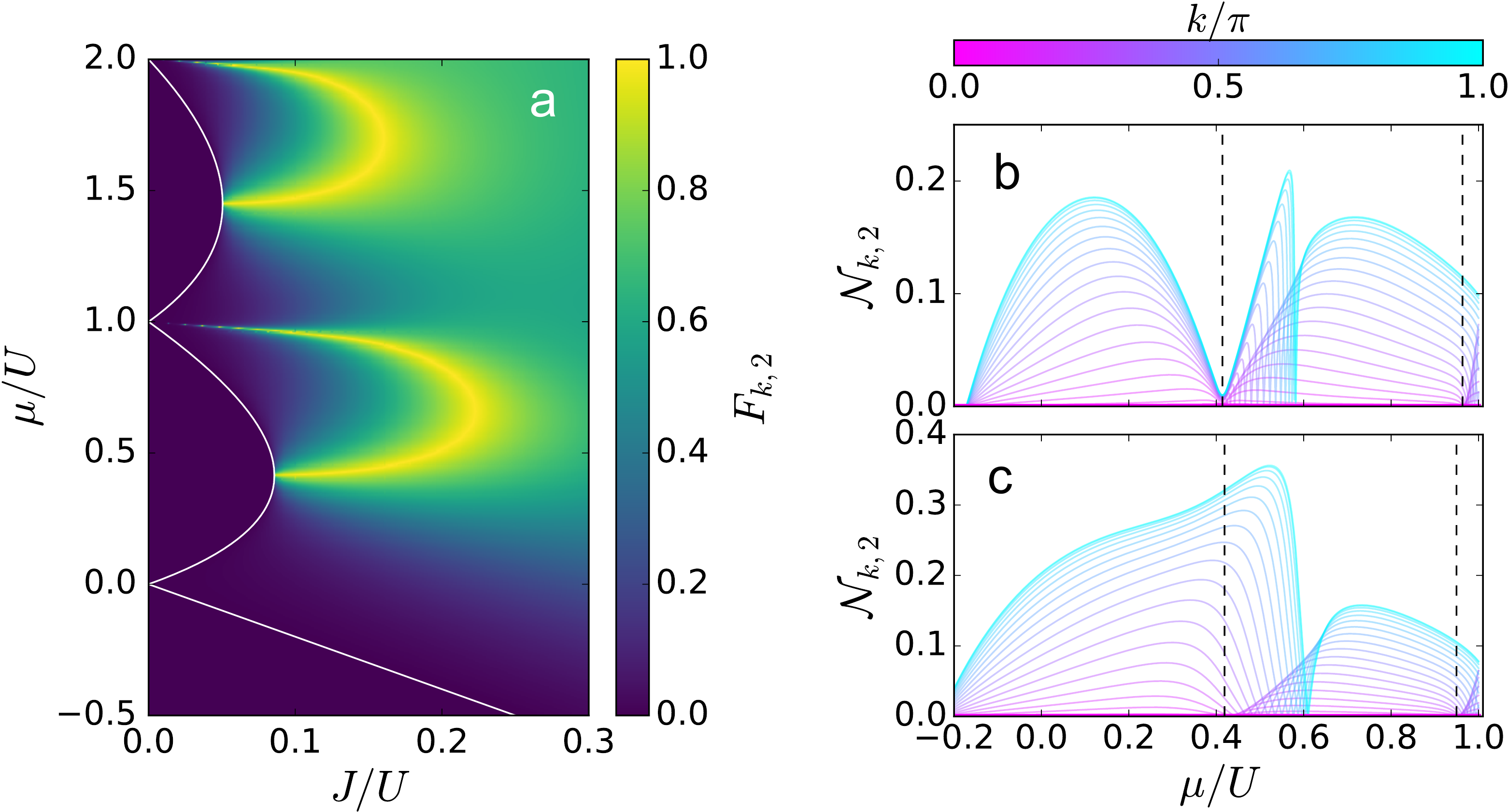}
\caption{(a) Flatness $F_{k, 2}$ for the first gapped mode ($\lambda = 2$) as
  a function of $J/U$ and $\mu/U$ for $k = \pi/100$; the bright
  yellow curves are the points where this mode corresponds to pure
  amplitude oscillation of the order parameter.  As a reference, the
  white lines indicate the Mott to superfluid phase boundaries. 
  (b) Density oscillation $\mathcal N_{k, 2}$ as a function of $\mu/J$ for fixed $J/U =
  0.0858$, corresponding to the tip of the lobe.  (c) As in (b) for
  $J/U=0.115$. Purple to light blue line color indicates $k$ varying
  from $0$ to $\pi$.  Vertical dashed lines highlight the zeros of
  $\mathcal N_{k, 2}$ at $k \approx 0$ (see (a)). }
\label{fig:Higgs}
\end{figure}

\emph{Pure amplitude (Higgs) mode.}
A long-standing debate has taken place about the conditions for the
existence of a gapped mode in the Bose-Hubbard model and its interpretation as a pure amplitude
oscillation of the superfluid order parameter \cite{Pekker2015}. Due to this
sought-after property, the first gapped mode is often referred to as
Higgs mode.
Within the linear approximation (see Eq.~(\ref{dphis})), pure
amplitude oscillations of the order paramenter $\psi_{i,\lambda}$ are
found when the imaginary part of $\delta \psi_{i,\lambda}$ vanishes,
namely when ${\cal I}_{\b k,\lambda} = \mathcal U_{\b k, \lambda} -
\mathcal V_{\b k, \lambda} = 0$.  Conversely, vanishing real part
(${\cal R}_{\b k,\lambda}=\mathcal U_{\b k, \lambda}+\mathcal V_{\b k,
  \lambda}=0$) corresponds to pure phase excitations of the order
parameter. 
To quantify the amplitude and phase
components of the oscillations of the order parameter in any mode
$\lambda$, it is useful to define the flatness parameter
\be F_{\b k,\lambda} =
\f{{\cal R}_{\b k,\lambda}
  - {\cal I}_{\b k,\lambda}}{{\cal R}_{\b k,\lambda} + {\cal I}_{\b k,\lambda}} \in
  [-1,1]\,.
  \ee
A positive flatness indicates a mode with dominant amplitude character
and a negative flatness indicates a mode with dominant phase
character.

In Fig.~\ref{fig:Higgs}(a), we show the flatness of the first gapped
mode ($\lambda=2$) at small momentum $\b k \approx 0$. This mode 
becomes purely amplitude-like ($F_{\b k,2}=1$) on the clear yellow
curve in the $(\mu/U,J/U)$ phase diagram.  The pure amplitude Higgs
mode emerges at the tip of each Mott lobe, where it is indeed expected
to exist, but quickly moves towards larger fillings as $J/U$
increases, and bends back towards $J/U\ra0$ and $\mu/U$ integer. This
behaviour confirms the expectations based on Fig.~\ref{fig:spectrum}
and related discussion. We stress that the initial and
final point of the curve ${\cal I}_{\b k,\lambda} =0$ are Lorentz
invariant points of the model.

Let us now define the density oscillations
\bea
\delta n_{i,\lambda} &=& 2 \mathcal N_{\b k, \lambda}
\cos(\b{k}\cdot \b{r}_i-\om_{\b k,\lambda} t), \label{dn}
\eea
with $ \mathcal N_{\b k, \lambda} = \sum_n \bar c_{n} n
(u^{(\lambda)}_{\b{k},n}+v^{(\lambda)}_{\b{k},n})$.
In correspondence of the particle-hole symmetry condition ${\cal I}_{\b k,\lambda} =0$,
the continuity equation  
yields $ \mathcal N_{\b k, \lambda}=0$, as confirmed by our calculations (see Fig.~\ref{fig:Higgs}(b,c)). This
identifies the pure amplitude character of a mode $\lambda$ with an exchange 
of particles between the condensate and the normal fraction. 

It is important to note that the pure amplitude character of the first
gapped mode is obtained on slightly different curves depending on the
momentum of the excitations. Moreover, the density response is
significant only in the short wavelength limit and it is suppressed
for $\b k \ra 0$ (see Fig.~\ref{fig:Higgs}(b,c)). These facts should
be taken into account when looking for the Higgs mode in possible
experiments \cite{Bissbort2011}.

\begin{figure}[!tbp]
  \includegraphics[width=1\columnwidth]{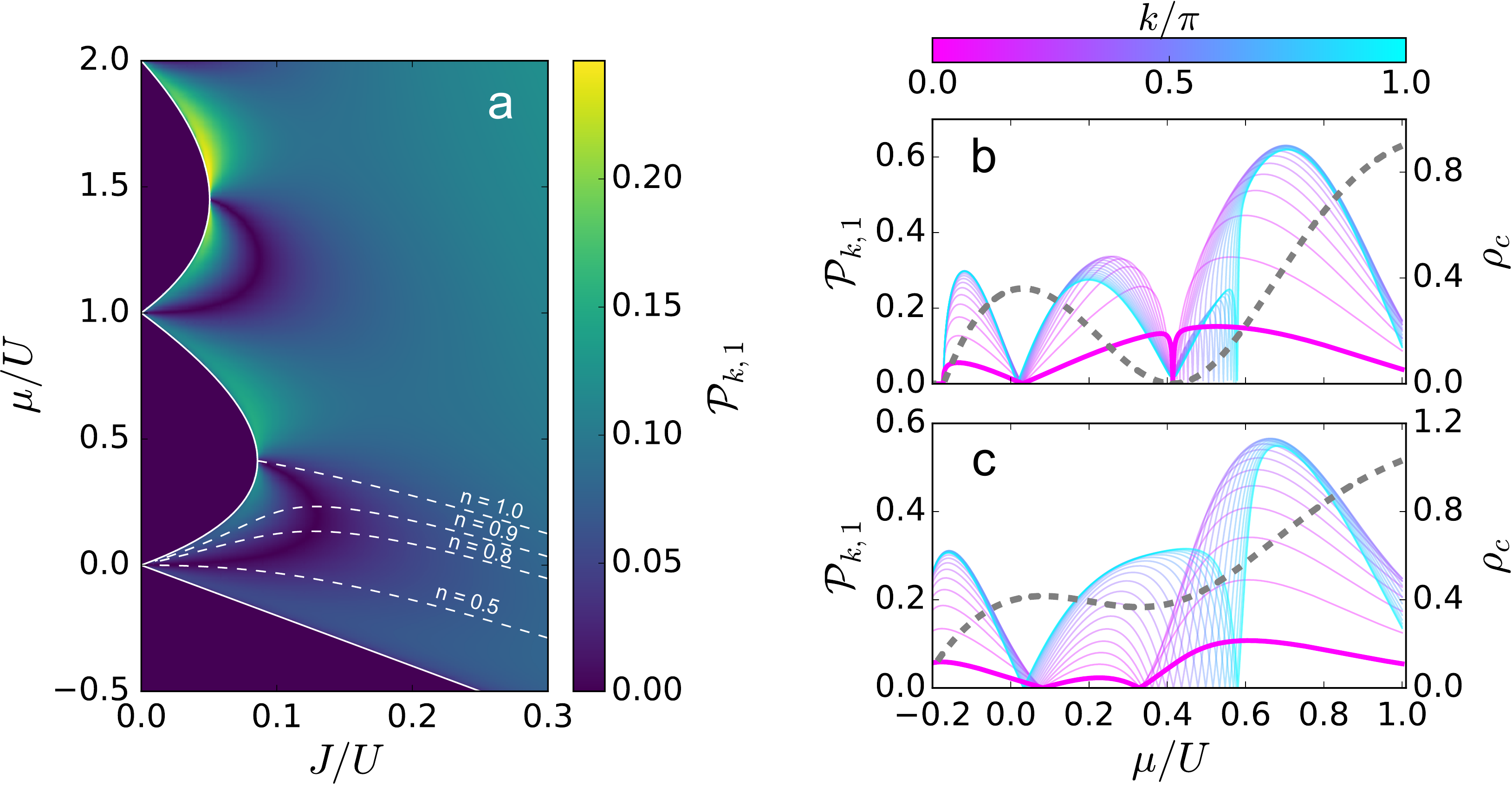}
\caption{(a) Amplitude of the condensate density fluctuations
  $\mathcal P_{k, 1}$ for the Goldstone mode ($\lambda = 1$) as a
  function of $J/U$ and $\mu/U$ for $k \approx \pi/100$; Constant
  density ${\bar n}= 0.5, 0.8, 0.9, 1$ contours (white lines).  (b)
  $\mathcal P_{k, 1}$ as a function of $\mu/J$ for fixed $J/U
  = 0.0858$, corresponding to the tip of the lobe. (c) As in (b) for
  $J/U=0.115$ (c).  Purple to light blue line color indicates $ k$
  vayring from $0$ to $\pi$. Grey dashed lines show the condensate
  density $\rho_c$ in the ground state. }
\label{fig:GOLD}
\end{figure}

\emph{Suppression of condensate density oscillations in the Goldstone
  mode.}  Particle and hole excitations of equal amplitude but
opposite sign (${\cal R}_{\b k, \lambda}=\mathcal U_{\b k,\lambda} +\mathcal V_{\b
  k,\lambda}= 0$) \footnote{Even though there exists a line $\mathcal R_{\b
    k, 1} = 0$, the property $\mathcal R_{\b k, 1}\ll \mathcal I_{\b
    k, 1}$ is fulfilled everywhere in the phase diagram for the linear
  part of the lowest branch of the spectrum, thus ensuring a phase
  character to the Goldstone mode.} directly imply vanishing
condensate density oscillations
\bea
\delta \rho_{c,i, \lambda} &=& \delta |\psi_{i, \lambda}|^2 = 2
\mathcal P_{\b k, \lambda}
\cos(\b{k}\cdot \b{r}_i-\om_{\b k, \lambda} t)\,,\label{drhoc}
\eea
with $\mathcal P_{\b k, \lambda} = \bar\psi \left( \mathcal U_{\b k,
  \lambda} + \mathcal V_{\b k, \lambda} \right)$.  Vanishing
condensate density oscillations are found on arc-shaped lines in the
phase diagram in the vicinity of, and in particular below, each Mott
lobe (dark blue curves in Fig.~\ref{fig:GOLD}(a)).  The suppression of
$\delta \rho_{c}$ for mode $\lambda = 1$ occurs for distinct
values $\b k$ on slightly different curves, which all lie above
half-integer filling and end in the vicinity of the tip of the lobe
(see Fig.~\ref{fig:GOLD}(b,c)).  Consistently, it will never be
possible to satisfy the condition ${\cal R}=0$ in the
weakly-interacting limit, where the Goldstone mode alone exhausts the
spectral function sum-rule $|\mathcal U_{\b k,1}|^2 - |\mathcal V_{\b
  k,1}|^2=1$.

In the superfluid hydrodynamic regime, the condensate
density oscillations of the Goldstone mode at low momenta couple only
to the density oscillations $\delta \rho_c = (\pa \rho_c / \pa n)_J
\delta n$.  This equality has been numerically verified by
independently calculating the oscillations $\delta \rho_c $, $\delta n
$ at $\b k\approx 0$ from Eqs.~(\ref{dn}, \ref{drhoc}) and the
quantity $ \pa \rho_c/\pa n $ in the ground state.  Hence, the
condensate density oscillations at $\b k \approx 0$ vanish in
correspondence of the maxima and the minima of the condensate density
at constant $J$ (see thick purple and dashed curves in
Fig.~\ref{fig:GOLD}(b,c)).

Remarkably, the suppression of condensate density
oscillations at $\b k \approx 0$ occurs at the boundary between particle and
hole superfluidity, usually defined as $(\pa \mu/\pa J)_n = 0$
\cite{Krutitsky2016}.
Indeed, the mean-field free energy per site $\Om$ depends on the
condensate density as $\Omega = -zJ \rho_c + \dots$, where $z=2d$ is
the coordination number in a hypercubic lattice and $d$ the number of
spatial dimensions.  Using the thermodynamic relations $\mu = (\pa
\Om/\pa n)_J$ and $\rho_c = -(1/z)(\pa \Om / \pa J)_n$, we obtain
$(\pa \mu / \pa J)_n = -z ( \pa \rho_c / \pa n )_J$.

Particularly interesting are the maximum of condensate
  density and the absence of condensate fluctuations found --
  essentially for all momenta -- on the lower branch of each ${\cal
    R}=0$ curve in Fig.~\ref{fig:GOLD}(a). This suggests the presence
  of a condensate that is extremely robust against thermal fluctuations for
  temperatures smaller than the Goldstone mode bandwidth and, as a
  possible consequence, an increase of the normal to superfluid
  critical temperature. This conjecture is supported by a qualitative
  comparison with quantum Monte Carlo results~\cite{Trotzky2010} 
  showing the critical temperature as a function
  of density at fixed $J/U$.  In Ref.~\cite{Trotzky2010}, for small $J$ and
  filling smaller than unity, a maximum of critical temperature is found
  above half-integer filling, in apparent agreement with the condition
  of particle-hole symmetry ${\cal R}=0$ found in this work. Moreover,
  the fact that the maximum of the critical temperature found
  in Ref.~\cite{Trotzky2010} is of the order of the hopping amplitude,
  namely, according to our calculation, smaller than the
  Goldstone mode bandwidth, validates an estimation of the critical
  temperature based on the thermal occupation of the Goldstone mode
  only. In this respect, the microscopic nature of the lowest-lying
  excitations and in particular their particle-hole symmetry seems to
  play a crucial role.
From a broader perspective, these findings may be relevant to understand 
the influence of Mott physics (Mottness) on the low-temperature phase diagram of cuprates
\cite{Paramekanti2001}.
Indeed, recent experiments have shown that -- among other effects \cite{Zaanen2015} -- 
at the optimal doping corresponding to the maximum of the 
superconducting dome, a transition from hole 
to particle transport \cite{Badoux2016} and a change in the charge 
transfer process \cite{Giannetti2011} occur.


\emph{Discussion.}  The unambiguous detection of particle-hole symmetry
  in the excitations of a strongly-interacting superfluid would
  require measurements able to independently resolve the amplitude and
  phase oscillations of the order parameter.
In other words, one needs to reconstruct the
single-particle Green's function in the laboratory. Pioneering experiments in this
directions have been performed in the early days of Bose-Einstein
condensation with two-pulse Bragg spectroscopy \cite{Brunello2000,
  Vogels2002}, where the particle and hole components of
Bogoliubov phonons have been resolved in real space after time-of-flight expansion. 
In the case of the Higgs mode, problems in applying this
technique arise due to its suppressed coupling with density
perturbations. 
One can also consider more sophisticated experimental techniques which are
presently being developed, namely ARPES-like schemes, as already
implemented with fermionic samples \cite{Dao2007,Stewart2008}, or
higher band Bragg spectroscopy, as already tested to detect the
correlations present in a bosonic Mott phase \cite{Fabbri2012}.
Proposals of lattice-assisted spectroscopy to emulate a STM 
(Scanning Tunneling Microscopy) in ultra-cold atomic
setups \cite{Kantian2015} and energy-resolved atomic scanning probes for
the density of states \cite{Gruss2016} have also been recently put forward.
In slightly different contexts, the Higgs
mode of a supersolid quantum gas has been created and detected by coupling a Bose-Einstein condensate to optical cavity modes \cite{Julian2017}. An even more speculative direction is given by the recent realization of Mott insulator states of light~\footnote{J. Simon, talk at BEC2017 Conference, San Feliu de Guixols, September 2017.}, which suggests the possibility of quantum simulating the Bose-Hubbard model in arrays of strongly-nonlinear optical or circuit-QED resonators~\cite{Hartmann2016, Noh2017, Lebreuilly2017}. In such optical systems, the full statistics of the quantum field is in fact directly accessible from a photoluminescence experiment~\cite{Carusotto2013}.

\emph{Conclusions.} In this Letter, we have discussed the particle-hole character
of the low-energy excitations in the strongly-interacting regime 
of the homogeneous Bose-Hubbard model. For the Goldstone mode, we have found 
that particle-hole symmetry induces a
suppression of condensate density oscillations below each Mott lobe
at the boundary between hole and particle superfluidity. As a consequence, 
we have conjectured an increase of the
normal to superfluid critical temperature.
Most remarkably, particle-hole symmetry also allows to identify the condition for the existence of the gapped pure-amplitude Higgs mode, which is found on a curve connecting the integer-density critical point (tip of the lobe) and the hard-core limit at half-integer density occurring between subsequent Mott lobes.

Our results rely on the fact that the structure of the excited modes is continuous 
across the phase transition and that the Mott phase is characterized 
by excitations with predominant particle or hole character.
We are confident that our conclusions are not affected by quantum 
fluctuations beyond the Gutzwiller approximation. Future directions might include the
development of a quantum theory for the excitations in order to account for 
the effect of quantum and thermal fluctuations, as well as possible decay processes.

\emph{Acknowledgements.} The authors would like to thank T. Comparin, C. Giannetti, L. Pollet and 
M. Punk for useful discussions. 
This work was supported by the EU-FET Proactive grant
AQuS, Project No.  640800 and by Provincia Autonoma di Trento. 


\end{document}